\documentstyle[prd,aps,psfig,eqsecnum]{revtex}  

\begin{document}
\tighten 

\def\ptitle#1{\wideabs{\maketitle
\abstract #1 \endabstract \pacs{Pacs numbers: 03.70.+k, 98.80.Cq}}}


%
%
%
%
  
\draft 

\title{Quantum corrections to critical phenomena in gravitational
collapse}  
\author{Patrick R Brady} 
\address{Theoretical Astrophysics,  California Institute of 
Technology, Pasadena, CA 91125} 
\author{Adrian C Ottewill} 
\address{Department of Mathematical Physics,  University College
Dublin,  Belfield,  Dublin 4, Ireland} 
\preprint{GRP-492}

\ptitle{ We investigate conformally coupled quantum matter fields on
spherically symmetric, continuously self-similar backgrounds.  By
exploiting the symmetry associated with the self-similarity the
general structure of the renormalized quantum stress-energy tensor can
be derived.  As an immediate application we consider a combination of
classical, and quantum perturbations about exactly critical collapse.
Generalizing the standard argument which explains the scaling law for
black hole mass, $M \propto |\eta-\eta^*|^\beta$, we demonstrate the
existence of a quantum mass gap when the classical critical exponent
satisfies $\beta \geq 0.5$.  When $\beta < 0.5$ our argument is
inconclusive; the semi-classical approximation breaks down in the
spacetime region of interest. }

\narrowtext

\section{Introduction} 

Choptuik~\cite{Choptuik_M:1993} demonstrated that, for suitably chosen
initial data, black holes of arbitrarily small mass can form in the
gravitational collapse of a massless scalar field.  Specifically, if the
strength of the initial data is characterized by some parameter
$\eta$, say, then there exists a critical value $\eta^*$ such that the
corresponding solutions to the Einstein-scalar field equations are
divided into three classes:
\begin{list}{}{\leftmargin=1.0em\itemsep=2pt\topsep=2pt\parsep=2pt}
\item[$\bullet$]\textit{Sub-critical solutions} have $\eta < \eta^*$;  the
collapsing matter eventually disperses leaving behind flat space.
\item[$\bullet$]{\textit{Critical solutions}} have $\eta \equiv
\eta^*$; they exhibit self-similar echoing in the neighborhood of a
central singularity.  The same echoing solution develops independent
of the shape of the initial data.
\item[$\bullet$]{\textit{Super-critical solutions}} have $\eta >
\eta^*$; the scalar field collapses to form a black hole.  The masses
of black holes which form in marginally super-critical evolutions obey
a scaling law such that
\begin{equation}
        M \propto |\eta-\eta^*|^\beta \label{eq:mass-scaling}
\end{equation}
where $\beta\approx 0.37$ is independent of the initial data.  
\end{list}
Since Choptuik's initial discovery, critical point behavior has been
studied in a variety of models for gravitational collapse
\cite{Evans_C:1994,Abrahams_A:1994b,Choptuik_M:1996,Hamade_R:1996}.   
Whenever black-hole formation turns on at infinitesimal mass,
precisely critical evolutions exhibit some form of self-similarity and
the black-hole mass scales as in Eq.~(\ref{eq:mass-scaling}) with a
model dependent exponent.

The properties of black holes are radically changed by quantum field
theory.  Since the work of Hawking~\cite{Hawking_S:1975} it is well
known that black holes formed by gravitational collapse will radiate
particles via quantum processes precisely as a black body with a
temperature, the Hawking temperature, proportional to the surface
gravity of the black hole.  For a Schwarzschild black hole the Hawking
temperature, $T_H$, is given by
\begin{equation}
  k T_H = { \hbar c^3 \over 8 \pi G M}
\end{equation}
where $M$ is the mass of the black hole.  Consequently, the black hole
radiates energy at a rate proportional to $M^{-2}$.  It follows that a
black hole of mass $M$ evaporates away by the Hawking process in
approximately $10^{-26} (M/1\;\mathrm{g})^3$ seconds; black holes of
very low mass, such as those formed in marginally super-critical
collapse, evaporate away almost instantaneously.
                
It would be of great interest to determine how the classical picture
of marginally supercritical collapse is modified by quantum gravity;
unfortunately this is beyond current techniques.  A more modest
program is to examine quantum fields in critical spacetimes, and from
this study to infer the semi-classical corrections to the classical
evolutions.  In this paper we undertake such an investigation.  We
focus attention on models of gravitational collapse in which
black-hole formation turns on at infinitesimal mass, {\em and} the
critical solution exhibits continuous self-similarity\footnote{By
continuous self-similarity, we mean that there exists a vector field
$\bbox{\xi}$ such that Eq.~(\protect\ref{eq:2}) is satisfied.}:
perfect fluid collapse and a class of Brans-Dicke models belong in
this category.  The continuous self-similarity allows us to infer a
great deal about the renormalized quantum stress-energy tensor for
conformally coupled fields in a precisely critical spacetime.
Moreover, generalizing the classical, perturbative treatments of Koike
{\it et al}~\cite{Koike_T:1995}, Maison~\cite{Maison_D:1996}, and
Gundlach~\cite{Gundlach_C:1997} to include semi-classical corrections,
we can infer the presence of a quantum mass-gap at the threshold of
black-hole formation when $\beta>0.5$.  (For perfect fluids with
pressure proportional to energy density, {\it i.e.} $p=k\rho$, $\beta$
increases monotonically from about $0.106$ when $k=0$ to $0.820$ when
$k=0.899$ passing through $\beta=0.5$ at $k\simeq 0.53$.) Our argument
is quite robust, requiring minimal assumptions about semi-classical
corrections to general relativity.  It is worth noting that a mass-gap
which originates with quantum effects will not be universal in
general.  This observation is a direct consequence of the non-locality
of the renormalized stress-energy tensor which carries information
about the initial data which led to the collapse.

We may contrast our approach with that of two related sets of work:
\begin{list}{}{}{\leftmargin=1.0em\itemsep=2pt\topsep=2pt\parsep=2pt}
\item[$\bullet$]
Ayal and Piran~\cite{Ayal_S:1997} have made a detailed numerical study
of scalar-field collapse in general relativity including a quantum
stress-energy tensor, inspired by two dimensional considerations, as a
source.  There are two differences with our approach: (1)~As Ayal and
Piran deal with scalar field collapse the critical spacetime in their
case contains only discrete self-similarity and not continous
self-similarity as we have assumed here.  (2)~The quantum
stress-energy tensor used by Ayal and Piran is not exactly conserved.
In contrast ours arises from a renormalised effective action and so is
conserved by constuction.
\item[$\bullet$]
Bose {\it et al}~\cite{Bose_S:1996,Peleg_Y:1997} have studied
semi-classical effects in gravitational collapse in the framework of
two-dimensional dilaton theories.  They have shown that quantum
effects lead to a mass-gap at the threshold of black-hole formation in
this theory.  
Unfortunately, it is not clear whether the critical solution in their
model exhibits any form of self-similarity, so it is difficult to
directly compare with the present work.
\end{list}
The paper is organized as follows.  In
section~\ref{sec:self-similarity} we introduce self-similarity in the
context of spherically symmetric spacetimes.  The purpose is to
highlight those features which are important in the subsequent
discussion of the renormalized stress-energy tensor (RSET).  This
discussion is presented in section~\ref{sec:RSET} which reviews the
properties of the conformal transformation law for the RSET of a
conformally invariant field.  This law contains anomalous terms
arising from the trace anomaly.  Self-similarity allows the metric of
the collapse spacetime to be written in a conformally stationary
form. The conformal transformation law for the RSET then allows
time-dependence of the RSET to be derived in these spacetimes.  In
section~\ref{sec:mass-scaling} we show how this information can be
incorporated into the standard classical, perturbative treatment to
include semi-classical corrections. From this analysis we infer that,
when semi-classical effects are accounted for, there is a mass gap at
the threshold of black-hole formation when the classical critical
exponent exceeds $\beta>0.5$.  We finish with a brief discussion of
the results.

\section{Self-similarity and spherical symmetry} 
\label{sec:self-similarity}

It is convenient to use a retarded coordinate $u$,  and to write the 
spherical line element as 
\begin{equation} 
ds^2 = e^{-2u} \bigl[ - G(u,\zeta) du^2 -2 H(u,\zeta) du d\zeta +  
        \zeta^2 d\Omega^2 \bigr] \; ,  
        \label{eq:ss-line-element} 
\end{equation} 
where $d\Omega^2 = d\theta^2 + \sin^2 \theta \, d\phi^2$.  Notice that
the radius of the two-spheres is given by
\begin{equation} 
        r(u,\zeta) = \zeta e^{-u} \; . 
\end{equation} 
Since these coordinates are unfamiliar it is worth elucidating a
couple of simple, but important, points.  The origin of the coordinate
$\zeta$ coincides with the symmetry origin, i.e. $r(u,0)=0$.  Surfaces
of constant $\zeta$ are timelike in a neighborhood of the origin;
generally this neighborhood does not extend over the entire patch
covered by the coordinates $(u,\zeta)$.  This is clearly demonstrated
by an example.  In Minkowski spacetime the metric functions are
$G(u,\zeta) = 1-2\zeta e^{-u}$ and $H(u,\zeta)=1$, so that $\zeta$
changes from timelike to spacelike when $\zeta=e^u/2$.  Nevertheless,
the metric is manifestly regular across this hypersurface.  Finally,
we note that $r\rightarrow \infty$ as $u\rightarrow -\infty$ on
surfaces of constant $\zeta$, while $r\rightarrow 0$ as $u\rightarrow
\infty$ along the same surfaces.
 
The spacetimes of interest below evolve from regular initial data, and
develop a singularity at $r=0$ as a result of gravitational
collapse of some matter field.  We normalize $u$ so that the
singularity is located at infinite coordinate time, and that the
proper time as measured by an observer at the origin is exponentially
related to $u$, that is $(\tau-\tau_s) \propto e^{-u}$ where $\tau_s$
is the proper time at the singularity.
 
Black-hole formation may be inferred from the existence of an apparent
horizon which expands to meet the event horizon at late times
(assuming that cosmic censorship holds).  Surfaces of constant radius
change from timelike to spacelike at the apparent horizon in spherical
spacetimes.  Thus, the normal to a surface of constant radius is null
at the apparent horizon, and the equation for an apparent horizon is
\begin{equation} 
        g^{\alpha\beta} \nabla_\alpha r \nabla_\beta r = 
        (G + 2 \zeta H) H^{-2} = 0 \; .
        \label{eq:apparent-horizon} 
\end{equation} 
 
Self-similar spacetimes are characterized by the existence of a vector
field $\bbox{\xi}$ such that
\begin{equation} 
        {\cal L}_{\bbox{\xi}} \bbox{g} = -2 \bbox{g} \; ,\label{eq:2}  
\end{equation} 
where $\bbox{g}$ is the metric tensor.  The above coordinates are well 
adapted to discuss self-similarity since the line element for a 
spherically symmetric, self-similar spacetime can always be written as 
in Eq.~(\ref{eq:ss-line-element}) with 
\begin{eqnarray} 
        G(u,\zeta) &&= G_{\rm ss} (\zeta) \; ,\\ 
        H(u,\zeta) &&= H_{\rm ss} (\zeta) \; , 
\end{eqnarray} 
and $\bbox{\xi} = \partial / \partial u$.  Explicitly, we have
\begin{equation} 
ds^2 = e^{-2u} \bigl[ - G_{\rm ss}(\zeta) du^2 -2 H_{\rm ss}(\zeta) du d\zeta +  
        \zeta^2 d\Omega^2 \bigr] \; ,  
        \label{eq:selfsim-line-element} 
\end{equation}
so that $g_{\mu\nu} = e^{-2u} \overline{g}_{\mu\nu}$ where $\bbox{\xi}
= \partial / \partial u$ is a Killing vector for the metric  
$\overline{g}_{\mu\nu}$.

\begin{figure}[h]
\vskip0.25in
\centerline{\psfig{file=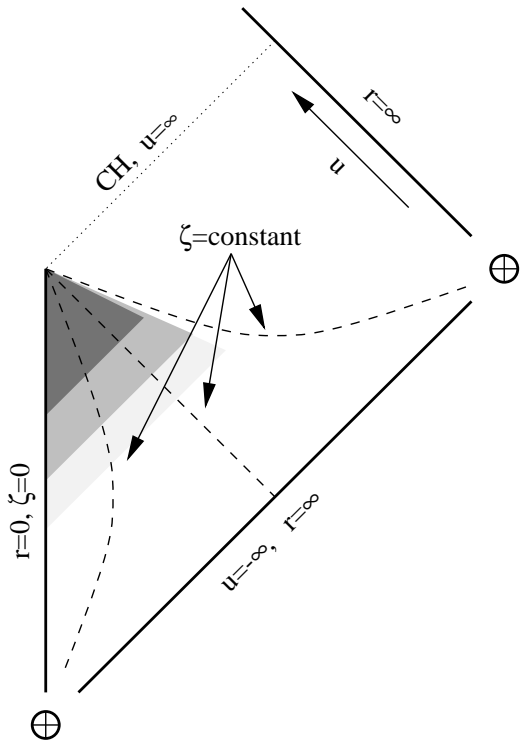,width=7cm,bbllx=160pt,bblly=180pt,bburx=500pt,bbury=600pt}}
\vskip0.25in
\caption{\label{fig:critical-spacetime}
Schematic representation of the spacetime of critical collapse.
Surfaces of constant $\zeta$ are indicated as dashed lines.  The
retarded time coordinate goes from $u=-\infty$ at past null infinity
to $u=\infty$ at the Cauchy horizon indicated by CH; a singularity
forms at $r=\zeta=0$, $u=\infty$.  The shaded region indicates the
asymptotic approach to self-similarity near the singularity.}
\end{figure}

In some studies of phase transitions in gravitational collapse,
continuous self-similarity is observed in near critical evolutions
when black-hole formation turns on at infinitesimal mass.  This is
schematically depicted in Fig.~\ref{fig:critical-spacetime} where the
spacetime diagram represents the collapse of critical ($\eta=\eta^*$)
initial data.  The shaded region indicates the asymptotic approach to
self similarity in the central region;   in {\em precisely} critical
evolutions this region extends all the way to the singularity.

\section{Semi-classical theory of critical collapse} 
 
The explicit computation of the renormalized stress-energy tensor for
quantum fields is beyond current techniques except in certain
exceptional circumstances with high symmetry.  Nonetheless,
significant progress can be made by exploiting the self-similarity of
critical spacetimes and a powerful tool which has been developed by
Page {\it et al}~\cite{Page_D:1982,Brown_M:1986}.
 
\subsection{Renormalized stress-energy tensor} \label{sec:RSET}

At the quantum level it is well known that renormalization breaks the
conformal invariance of a classically conformally invariant theory.
This is manifested in the existence of the trace anomaly.  As a result
the renormalized stress-energy tensor does not simply scale
under conformal transformation but also acquires geometrical
corrections.  Quite generally, Page has shown that the
RSET for conformally coupled fields in a spacetime $({\cal M}, g)$ can
be obtained from the RSET in the conformally related spacetime
$({\cal M},\overline{g})$, where $g_{\mu\nu} = e^{-2\omega}
\overline{g}_{\mu\nu}$, by the following transformation rule:
\begin{eqnarray} 
\label{eq:transformation} 
        \left< T_{\mu}{}^\nu \right> &=& e^{4\omega} \overline{\left< 
 T{_\mu}^\nu \right> } 
 + 8 \alpha e^{4\omega}  \left[ (\omega \overline{C}_{\alpha\mu\beta}{}^{\nu} 
 )^{:\alpha\beta} +  
  {\textstyle {1 \over 2}} \omega \overline{R}^{\alpha\beta}\overline{C}_{\alpha\mu\beta}{}^{\nu} 
  \right]   \nonumber \\  
&& \hspace{-0.3in} - \beta  \bigl[  (2H_{\mu}{}^\nu - 4 R^{\alpha}{}_{\beta}
C_{\alpha\mu}{}^{\beta\nu}) - e^{4\omega} (2\overline{H}_{\mu}{}^\nu -  4 
 \overline{R}^{\alpha}{}_{\beta}\overline{C}_{\alpha\mu}{}^{\beta\nu}) 
\bigr] \nonumber \\
&& \hspace{-0.3in} - {\textstyle {1 \over 6}}
\gamma\left[I_{\mu}{}^\nu - e^{4\omega}  \overline{I}_{\mu}{}^\nu
\right]  \; .
\end{eqnarray} 
Here a colon denotes covariant differentiation with the natural
connection for the metric $\overline{g}_{\mu\nu}$, 
\begin{equation} 
\label{eq:H_tensor} 
 {H}^{\mu\nu} = - R^{\mu}{}_{\alpha} R^{\alpha\nu} + {\textstyle {2 \over 3}} R 
 R^{\mu\nu} + ({\textstyle {1 \over 2}} R_{\alpha\beta}R^{\alpha\beta}  
- {\textstyle{1  \over 4}} R^2) g^{\mu\nu} \; ,
\end{equation} 
and 
\begin{eqnarray} 
\label{eq:I_tensor} 
  {I}^{\mu\nu}  &=& {1 \over \sqrt{g}} {\delta \ \over \delta g_{\mu\nu}}  
    \int \sqrt{g}\, {\rm d}^4 x \> R^2     \nonumber\\ 
&=& 2R^{;\mu\nu} - 2R R^{\mu\nu} + 
  ({\textstyle {1 \over 2}} R^2 - 2 R_{;\alpha}{}^{;\alpha}) g^{\mu\nu}  . 
\end{eqnarray} 
Here,  $\left< T{_\mu}^\nu \right>$ and $\overline{\left< T{_\mu}^\nu 
  \right>}$ denote the RSET in some state on $({\cal M}, g)$  and in the 
conformally related state on $({\cal M},\overline{g})$. 
The coefficients $\alpha$, $\beta$ and $\gamma$ depend on the spin of 
the field; if $h_s$ denotes the number of helicity states for fields of 
spin $s$ 
then 
\begin{mathletters} 
\begin{eqnarray} 
\label{eq:anomaly_coefficients} 
\alpha &=& \bigl[12 h_0 + 18 h_{1 \over 2} + 72
  h_1\bigr]/(2^9 45\pi^2) \\ 
\beta &=& \bigl[-4 h_0  -11 h_{1 \over 2} -124
  h_1\bigr]/(2^9 45\pi^2) \\ 
\gamma &=& \bigl[8 h_0 + 12 h_{1 \over 2}+ (48 \mbox{\
    or} -72) h_1\bigr]/(2^9 45\pi^2) . 
\end{eqnarray} 
\end{mathletters} 
The ambiguity in the coefficient of $h_1$ arises from the choice of 
renormalization method but is irrelevant to our discussion. 
  
For our purposes, it is convenient to rewrite the first term using 
the identity 
\begin{eqnarray} 
 (\omega &&\overline{C}_{\alpha\mu\beta}{}^{\nu} )^{:\alpha\beta} +  
 {\textstyle {1 \over 2}}\omega
 \overline{R}^{\alpha\beta}\overline{C}_{\alpha\mu\beta}{}^{\nu} = \nonumber \\ 
    &&\omega \overline{B}_{\mu}{}^\nu   
   + \omega^{:\beta} \overline{C}_{\alpha\mu\beta}{}^{\nu:\alpha} 
  +  \omega^{:\alpha} \overline{C}_{\alpha\mu\beta}{}^{\nu:\beta} 
  + \omega^{:\alpha\beta} \overline{C}_{\alpha\mu\beta}{}^{\nu}   
\end{eqnarray} 
where the Bach tensor $B^{\mu\nu}$ is defined as 
\begin{eqnarray} 
\label{eq:B_tensor} 
  {B}^{\mu\nu}  &=&  {1 \over \sqrt{g}} {\delta \ \over \delta g_{\mu\nu}}  
    \int \sqrt{g}\, {\rm d}^4 x \> 
    {\textstyle {1 \over 4}}C_{\alpha\beta\gamma\delta}
	C^{\alpha\beta\gamma\delta} \nonumber\\ 
&=& {C}_{\alpha}{}^\mu{}_\beta{}^{\nu;\alpha\beta} +  
 {\textstyle {1 \over 2}} R^{\alpha\beta}C_{\alpha}{}^\mu{}_\beta{}^{\nu} . 
\end{eqnarray} 
Furthermore, since \cite{Hawking_SW:1973} 
\begin{equation} 
 R^{\alpha\beta}{}_{\gamma\delta} = e^{4\omega} \left[  
      \overline{R}^{\alpha\beta}{}_{\gamma\delta} + 
         \mbox{terms involving $\omega_{:\rho}$} \right] 
\end{equation} 
we find  
\begin{eqnarray} 
\label{eq:general_law} 
        \left< T_{\mu}{}^\nu \right> &=& e^{4\omega} \overline{\left< 
 T{_\mu}^\nu \right> } 
 + 8 \alpha \omega e^{4\omega} \overline{B}_{\mu}{}^\nu \nonumber \\ 
&& +  e^{4\omega} \left[ \mbox{terms involving $\omega_{:\rho}$} \right] . 
\end{eqnarray}

There is an ambiguity in the RSET relating to the $\omega$
contribution, that is, the {\em logarithm} of the conformal factor in
Eq.\ (\ref{eq:transformation}) that we must now discuss.  There is an
arbitrary renormalization scale hidden in the logarithm so that a
constant conformal transformation, which corresponds simply to a
change in length scale, changes the RSET by the addition of a multiple
of the Bach tensor.  (Since the Bach tensor arises from a conformally
invariant action it is traceless, so this ambiguity does not effect
the trace anomaly.) In the next subsection we will see that this
merely corresponds to a choice of renormalization point for the
coupling coefficients in a generalized Einstein action.
  
A direct application of Eq.\ (\ref{eq:general_law}) to the
self-similar spacetimes of Eq.\ (\ref{eq:selfsim-line-element}) with
$e^{-\omega}= e^{-u}$ determines the $u$ dependence of the RSET in the
physical spacetime.  Schematically, we can write
\begin{eqnarray}
\label{eq:selfsim_law} 
        \left< T_{\mu}{}^\nu \right> = e^{4u} \overline{\left< 
 T{_\mu}^\nu \right>} (\zeta)  
 + 8 \alpha u e^{4u} \overline{B}_{\mu}{}^\nu (\zeta) 
 +  e^{4u} \overline{S}_{\mu}{}^\nu (\zeta),  
\end{eqnarray} 
where $\overline{S}_{\mu}{}^\nu (\zeta)$ denotes a tensor constructed
from the geometry of $({\cal M},\overline{g})$ and
$\omega_{:\mu} = \delta^u_\mu$ which cannot depend on $u$ since $\bbox{\xi} =
\partial / \partial u$ is a Killing vector of $({\cal
M},\overline{g})$.  The state dependence is carried by the RSET
$\overline{\left< T{_\mu}^\nu \right>}(\zeta)$ computed in the
conformal spacetime; it is independent of $u$ since $\bbox{\xi}$ is a
Killing vector in this spacetime, and we expect the quantum states of
interest to respect this symmetry.
  
\subsection{Semi-classical equations} 
In general, quantum field theory in curved spacetime is only 
renormalizable (at one-loop) when viewed as part of a general theory 
of the gravitational field with a low-energy effective action of the 
form 
\begin{equation} 
        I = \int_{\cal M} \left({1 \over 16\pi G} (R -2  \Lambda) + 
        { {a \over 4}} C_{\alpha\beta\gamma\delta} 
        C^{\alpha\beta\gamma\delta} + b 
        R^2\right) \; . 
\end{equation} 
The coupling constants $\Lambda$, $a$ and $b$ of this effective theory
must be measured.  Therefore the ambiguity associated with the
regularization scale is a manifestation of our lack of knowledge of
physics at Planck scales -- the boundary of validity of any effective
theory based on an expansion of the gravitational action in powers of
curvature.  The generalized (semi-classical) Einstein equations can be
written as
\begin{equation} 
        G^{\mu\nu} + \Lambda g^{\mu\nu} = 
  8\pi G \left[ T^{\mu\nu} +  \epsilon\left(\left< T^{\mu\nu} 
                \right> + 2 a B^{\mu\nu} + 2 b I^{\mu\nu} \right)\right] 
        \label{eq:semiclassical} 
\end{equation} 
where $\epsilon$ is a counting parameter which is unity if
semi-classical effects are included and zero otherwise.
  
The classical, critical solution corresponds to a self-similar
solution to these equations with $\epsilon=0$ and $\Lambda=0$.  We
wish to consider perturbations to such solutions which originate with
small deviations from critical initial data in the presence of quantum
matter.  A non-zero cosmological constant would change the value of
$\eta$ at the critical point, but the asymptotic solution should be
unchanged provided $1/\sqrt{\Lambda}$ is larger than the initial
matter configuration.  For this reason we assume that $\Lambda$
remains zero.  Thus, we look for solutions of these generalized
equations of the form
\begin{eqnarray} 
        G(\zeta, u) &&= G_{\rm ss}(\zeta) - (\eta-\eta^*) g_c(\zeta) 
        e^{\omega_c u} + \epsilon g_q(\zeta) e^{\omega_q u} \label{eq:G}\\ 
        H(\zeta, u) &&= H_{\rm ss}(\zeta) - (\eta-\eta^*) h_c(\zeta) 
        e^{\omega_c u} + \epsilon h_q(\zeta) e^{\omega_q u} \label{eq:H}
        \; .
\end{eqnarray}
It is unnecessary to consider changes in $g_{\theta\theta}$ since they can
always be removed by a first order coordinate transformation.  The
value of $\omega_c$ is determined by solving a boundary value problem
for the classical perturbations of the self-similar solution; this has
been done by several
authors~\cite{Maison_D:1996,Koike_T:1995,Gundlach_C:1997}.  However,
$\omega_q=2$ is easily determined by computing the Einstein tensor to
linear order in $\hbar$ for the line-element in
Eq.\ (\ref{eq:ss-line-element}) [with $G(\zeta,u)$ and $H(\zeta,u)$
determined by Eqs.\ (\ref{eq:G}) and (\ref{eq:H}) respectively], and
comparing the $u$ dependence with that of the RSET in
Eq.\ (\ref{eq:semiclassical}) as determined by Eq.\ (\ref{eq:selfsim_law}).
 
\subsection{Modified mass scaling}\label{sec:mass-scaling}

We can now consider the modified scaling relation for black hole mass.
In the self-similar spacetimes corresponding to the critical point of
gravitational collapse $G_{\rm ss} + 2 H_{\rm ss} \zeta > 0$
everywhere, i.e.,  there is no apparent horizon.
 
Substituting the perturbed quantities into
Eq.~(\ref{eq:apparent-horizon}), the apparent horizon is located at
$(\zeta_h, u_h)$ such that
\begin{eqnarray} 
        F(\zeta_h,u_h) &=&  
        G_{\rm ss} + 2 H_{\rm ss} \zeta_h - (\eta-\eta^*) (g_c+2h_c\zeta_h) 
        e^{\omega_c u_h} \nonumber \\
        && \hspace{0.2in} + \epsilon (g_q+2h_q\zeta_h) 
        e^{\omega_q u_h} = 0 \; .  \label{eq:bhrad1}
\end{eqnarray} 
Now, the radius of the apparent horizon is related to $(u_h,\zeta_h)$
by $R_h = e^{-u_h} \zeta_h$ so that Eq.~(\ref{eq:bhrad1}) can be
rewritten as 
\begin{eqnarray} 
        F(R_h,\zeta_h) &=& G_{\rm ss} + 2 H_{\rm ss} \zeta_h -  
        (\eta-\eta^*) (g_c+2h_c\zeta_h) 
        \zeta_h^{\omega_c} R^{-\omega_c} \nonumber \\
        && \hspace{0.2in} + \epsilon (g_q+2h_q\zeta_h) 
        \zeta_h^{\omega_q} R^{-\omega_q} = 0 \; . \label{eq:bhrad2}
\end{eqnarray} 
The classical limit ($\epsilon =0$) has been explored by other authors
who have argued that the observed scaling relation for black hole mass
is determined by solving Eq.~(\ref{eq:bhrad2}) for $R_h$ in this
limit~\cite{Koike_T:1995,Maison_D:1996}.  Thus, one arrives at the
relation
\begin{equation} 
        R_h \propto (\eta-\eta^*)^{1/\omega_c} \; .
\end{equation} 
For perfect fluids with pressure proportional to energy density, 
i.e., $p=k\rho$, the classical parameter $\omega_c$ decreases
monotonically from about $9.46$ when $k=0$ to $1.22$ when $k=0.899$
passing through $\omega_c=2$ at $k\simeq
0.53$~\cite{Koike_T:1995,Maison_D:1996}.

As the mass of the black hole which forms in marginally super-critical
collapse approaches the Planck mass, quantum effects will become
significant.  Moreover, it is reasonable to expect that quantum matter
will compete with gravitational collapse eventually averting formation
of a black hole for some $\eta_q$ sufficiently close to $\eta^*$.  The
conclusions that can be drawn from our analysis depend strongly on the
relative magnitudes of $\omega_c$ and $\omega_q$, therefore we break
the discussion into two cases.

\begin{figure} 
\centerline{\psfig{file=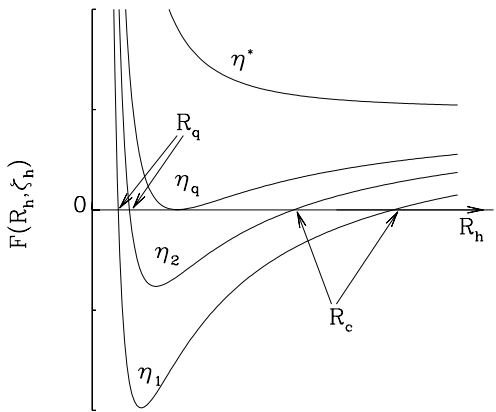,width=7cm}} 
\caption{\label{fig:mass-gap}
The horizon location is determined by the roots of the function
$F(R_h,\zeta_h)$ in Eq.~(\protect\ref{eq:bhrad2}).  We show here a
schematic representation for several values of $\eta$ which determine
deviations from classically critical initial data, and
$\omega_c<\omega_q$.  For sufficiently large $\eta$ classical collapse
takes hold and a black hole forms at $R_c$.  The function has a
minimum,  however,  and another root $R_q$ exists.  As $\eta$ is tuned to a
critical value $\eta_q$ the two roots coincide.  When $\eta<\eta_q$ no black
hole forms.  Thus a mass gap exists at the threshold of black-hole
formation in semi-classical collapse.}
\end{figure}

(i) When $\omega_c<\omega_q$ the function $F(R_h,\zeta_h)$ is
represented schematically in Fig.~\ref{fig:mass-gap}.  It is
approximately constant during the self-similar phase of the evolution.
For sufficiently large $\eta$, classical gravitational collapse takes
hold and a black hole forms at $R_c$.  Note that the function has a
minimum at smaller $R_h$, and a second root at $R_q$.  As $\eta$
decreases $R_c\rightarrow R_q$ until the roots coincide at some
critical value of the parameter $\eta_q$.  When $\eta<\eta_q$, no
black hole forms.  Thus, we can infer a mass-gap at the threshold of
black-hole formation in semi-classical collapse.
 
\begin{figure} 
\centerline{\psfig{file=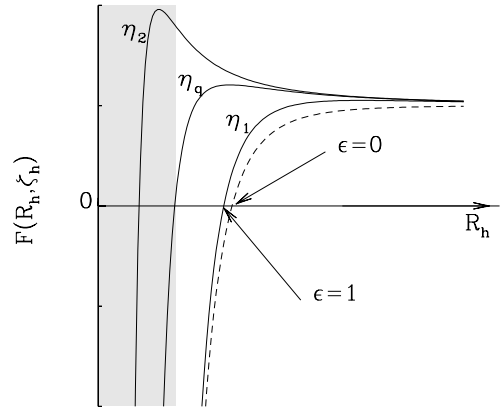,width=7cm}} 
\caption{\label{fig:inconclusive}
The horizon location is determined by the roots of the function
$F(R_h,\zeta_h)$ in Eq.~(\protect\ref{eq:bhrad2}).  When $\omega_c\geq
\omega_q$ and $\eta\rightarrow \eta^*$ the function has a local maximum,
however the classical terms always dominate as $R_h\rightarrow 0$.
The dashed line is $F(R_h,\zeta_h)$ in the absence of quantum
corrections when $\eta=\eta_1$.  By assumption, quantum corrections
decrease the size of the black hole as indicated by the slight
decrease in the root when $\epsilon=1$ and $\eta=\eta_1$.  When
$\eta=\eta_q$ the black-hole horizon lies at the boundary of Planckian
curvature and we must appeal to quantum gravity to understand the
quantum corrections to the near critical evolutions.}
\end{figure}

(ii) When $\omega_c\geq\omega_q$ we can say less about the critical
point.  Figure~\ref{fig:inconclusive} shows $F(R_h,\zeta_h)$ in this
circumstance for the two cases $\epsilon=0$ and $\epsilon=1$.  As
$\eta\rightarrow \eta^*$ semi-classical effects have a significant
effect causing the radius of the apparent horizon to be reduced
compared to the purely classical result; there is only a single root
of Eq.~(\ref{eq:bhrad2}).  Once again, there is a critical value
$\eta_q$ of the parameter which marks the point when the apparent
horizon radius corresponds to the boundary at which curvatures reach
Planck scales and we can no longer trust semi-classical calculations.
Quantum gravity, or at least a better approximation to it, is
needed to properly determine the critical point behavior.

\section{Discussion} 

The conclusion arrived at here is not rigorous.  We do not fully
understand quantum gravity, or how to fully incorporate semi-classical
effects into gravity.  Nonetheless we have been able to make some
progress in understanding semi-classical effects in critical
spacetimes by studying the structure of the renormalized stress-energy
tensor for conformally coupled fields in the critical background
spacetime.  By modifying the perturbative arguments which are used to
obtain the critical exponent observed in classical collapse we have
been able to infer a mass gap at the threshold of black-hole formation
in semi-classical theory.  This conclusion relies heavily on the
assumption that quantum matter tends to oppose black-hole formation.
The validity of this assumption can not be addressed without a
complete calculation of the renormalized stress-energy tensor in the
dynamical spacetimes of near critical collapse.  Such a computation
would be very difficult requiring the development of new techniques.

\acknowledgements

We would like to thank Leonard Parker for discussions. This work was
supported in part by an International Collaboration Grant from
Forbairt, and NSF grant AST-9417371.  P.R.B. is grateful to the
Sherman Fairchild Foundation for financial support.

\end{document}